\documentclass[10pt,conference]{IEEEtran}
\IEEEoverridecommandlockouts

\usepackage{amsmath,amssymb,amsfonts}
\usepackage{algorithmic}
\usepackage{graphicx}
\usepackage{textcomp}
\usepackage{xcolor}
\usepackage{capt-of}
\usepackage{minted}
\usepackage[ruled,linesnumbered]{algorithm2e}
\usepackage{import}
\usepackage{tikz}
\usepackage{moresize}
\usepackage{enumerate}
\usepackage{listings}
\usepackage{booktabs}
\usepackage{multirow}
\usepackage{colortbl}
\usepackage{array}
\usepackage{cite}
\def\BibTeX{{\rm B\kern-.05em{\sc i\kern-.025em b}\kern-.08em
    T\kern-.1667em\lower.7ex\hbox{E}\kern-.125emX}}

\begin{document}

\title{Evaluating Large Language Models for Functional and Maintainable Code in Industrial Settings: \\ A Case Study at ASML}

\author{
\IEEEauthorblockN{Yash Mundhra}
\IEEEauthorblockA{\textit{Delft University of Technology}\\
Delft, Netherlands \\
y.mundhra@student.tudelft.nl}
\and
\IEEEauthorblockN{Max Valk}
\IEEEauthorblockA{\textit{ASML}\\
Veldhoven, Netherlands \\
max.valk@asml.com}
\and
\IEEEauthorblockN{Maliheh Izadi}
\IEEEauthorblockA{\textit{Delft University of Technology}\\
Delft, Netherlands \\
m.izadi@tudelft.nl}
}

\maketitle

\begin{abstract}
Large language models have shown impressive performance in various domains, including code generation across diverse open-source domains. However, their applicability in proprietary industrial settings, where domain-specific constraints and code interdependencies are prevalent, remains largely unexplored. We present a case study conducted in collaboration with the leveling department at ASML to investigate the performance of LLMs in generating functional, maintainable code within a closed, highly specialized software environment.

We developed an evaluation framework tailored to ASML's proprietary codebase and introduced a new benchmark. Additionally, we proposed a new evaluation metric, \textit{build@k}, to assess whether LLM-generated code successfully compiles and integrates within real industrial repositories. We investigate various prompting techniques, compare the performance of generic and code-specific LLMs, and examine the impact of model size on code generation capabilities, using both match-based and execution-based metrics.
The findings reveal that prompting techniques and model size have a significant impact on output quality, with few-shot and chain-of-thought prompting yielding the highest build success rates. The difference in performance between the code-specific LLMs and generic LLMs was less pronounced and varied substantially across different model families.
\end{abstract}


\section{Introduction}
The rise of large language models (LLMs) has significantly influenced software engineering practices, with tools like GitHub Copilot and ChatGPT already supporting developers in writing, refactoring, and understanding code~\cite{hou_large_2024}. Despite their promising capabilities, most LLMs are trained on public, open-source data and evaluated on standardized benchmarks that focus on generating straightforward, standalone code snippets~\cite{jiang_survey_2024,fan_large_2023}. As a result, their effectiveness in proprietary, domain-specific industrial settings remains unclear. In practice, industrial software systems often involve legacy components, tightly coupled modules, specialized APIs, and unique naming conventions~\cite{joel_survey_2024,zhang_repocoder_2023}. All of which pose significant challenges for LLM-based code generation.

In this work, we investigate whether and how LLMs can be effectively applied in real-world industrial code bases that lie outside their training dataset. Specifically, we explore the applicability of LLM-based code generation in the ASML leveling department, where code operates within a complex, layered architecture. We examine whether LLMs can generate interdependent, buildable, and functional code in a proprietary repository with no or limited prior exposure to its domain-specific terminology or structure. 

Existing literature has largely focused on function-level or benchmark tasks with clear input-output mappings~\cite{koohestani2025benchmarking}. Few studies have explored repository-level code generation in closed industrial environments~\cite{izadi2024language,de2024transformer}. This gap leaves open questions around the feasibility of LLMs for code generation in an industrial setting. Aspects such as prompt engineering, model specialization, model size, and functional correctness have been studied very minimally in repository-level settings.

We address these open questions through an in-depth empirical study conducted at ASML. By evaluating LLMs across prompting strategies and model configurations, and by introducing a novel evaluation metric (build@k), we provide practical insights into the challenges and opportunities of adopting LLMs for domain-specific software generation. 

Our contributions are as follows.
\begin{itemize}
        \item Our study demonstrates how LLMs perform when tasked with generating code in a closed, domain-specific repository within a proprietary domain.

        \item We created a custom benchmark using ASML's internal code to evaluate the generation capabilities of LLMs.
 
        \item We propose a novel metric, \textit{build@k}, to evaluate whether generated code successfully compiles and builds, providing a more realistic measure of usefulness than traditional similarity-based metrics.

        \item We compare zero-shot, few-shot, and chain-of-thought (CoT) prompting techniques and find that the latter two significantly outperform zero-shot in this domain.

        \item We show that code-specific LLMs generally outperform generic ones, though the gap varies across model families.

        \item Lastly, our research highlights that larger models tend to perform better, but gains diminish beyond a certain size (around 14B parameters), suggesting diminishing returns on scaling.
\end{itemize}

\section{Background and Related Work}
The pivotal research by Hindle et al.~\cite{hindle_naturalness_2012} proved that software, although theoretically complex, is indeed predictable through statistical modeling. It is therefore not surprising that code completion and code generation have become some of the most thoroughly researched applications of LLMs in software engineering, leveraging this predictable nature to generate effective code recommendations~\cite{izadi2022codefill,van2024investigating}. Additionally, a survey conducted by Hou et al.~\cite{hou_large_2024} emphasized that current research predominantly focuses on applying LLMs during the software development phase of the engineering life cycle.

Within this growing field of research, a distinction can be made between \textit{general-purpose} LLMs and \textit{code-specific} LLMs. General-purpose LLMs, such as GPT-4~\cite{openai_gpt-4_2024} and LLaMa 3~\cite{grattafiori_llama_2024}, are trained on a vast and diverse corpus of text, including web data, code, documents, and news articles, which provides them with a broad knowledge base. These models have demonstrated impressive performance in various software engineering tasks, including code writing, understanding, and reasoning. In contrast, code-specific LLMs are typically trained on a massive corpus of programming data or fine-tuned from a general-purpose LLM using a large amount of programming data~\cite{jiang_survey_2024}. By focusing on programming-related tasks and challenges, these code-specific models have achieved even better performance than generic LLMs when it comes to generating functionally correct code.

Since the release of CodeX, a decoder-only language model fine-tuned for programming tasks,~\cite{chen_evaluating_2021}, research on LLMs for code generation has accelerated. The introduction of the HumanEval benchmark, created to evaluate functional correctness from docstrings, played a key role in this surge, becoming a standard for assessing model performance in code synthesis. In 2024, Zhao et al.~\cite{team_codegemma_2024} released CodeGemma, a decoder-only model built on Google's Gemma architecture, trained on 500 billion tokens of primarily code data, achieving state-of-the-art performance in code generation and completion tasks.

That same year, the DeepSeek-AI team launched DeepSeek-Coder-V2~\cite{deepseek-ai_deepseek-coder-v2_2024}, an open-source model pre-trained on 6 trillion additional tokens, enhancing its coding and mathematical reasoning capabilities while maintaining performance in general language tasks. The model outperformed all open-source counterparts and matched leading closed-source models like GPT-4 Turbo. Shortly after, Hui et al.~\cite{hui2024qwen2} released Qwen2.5-Coder, a decoder-only model from Alibaba, which significantly improved upon its predecessor by being pre-trained on over 5.5 trillion tokens of code-centric data.

While LLMs show strong capabilities in code generation, benchmarks like HumanEval focus on simple tasks, such as generating standalone functions or statements. Software development, however, involves complex dependencies and interdependent code units~\cite{du_evaluating_2024}. Jimenez et al.~\cite{jimenez_swe-bench_2024} assessed LLMs in realistic settings, where models resolved issues in GitHub repositories, with the best model, Claude 2, resolving only 1.96\% of issues. This highlights the need for improved domain-specific code generation, a field still largely underexplored.

To tackle more challenging coding scenarios, Du et al.~\cite{du_evaluating_2024} introduced ClassEval, which involves generating classes based on descriptions, test suites, and benchmark solutions, using a class skeleton as a blueprint. Despite its contributions, ClassEval treats classes as isolated units, relying only on common libraries likely included in LLM training data. Yu et al.~\cite{yu_codereval_2024} proposed CoderEval, evaluating LLMs on pragmatic code generation using non-standalone functions from real-world projects. Yet, CoderEval is limited to line- and function-level tasks. Deshpande et al.~\cite{deshpande_class-level_2024} introduced RepoClassBench, assessing LLMs on generating non-standalone classes within a repository's context, offering a more realistic reflection of real-world scenarios. Nevertheless, despite these advances, research on industrial-scale code generation remains relatively limited and underexplored.

When using LLMs, the quality of the prompt plays a crucial role in shaping the output they generate~\cite{marvin_prompt_2024}. In software engineering tasks, three popular prompting techniques are used the most: zero-shot, few-shot, and chain-of-thought prompting~\cite{hou_large_2024}. Zero-shot prompting asks the model to tackle a task without any examples, relying on its existing knowledge to figure things out. Few-shot prompting provides a few examples in the prompt to help the model understand the task better, leading to more accurate results. Chain-of-thought prompting takes it a step further by encouraging the model to break down complex tasks into logical steps, which is great for handling intricate problems that require sequential thinking. While these techniques have proven effective in various scenarios, their impact on generating repository-level domain-specific code is not yet well-studied.

\section{Problem and Industrial Context}
\begin{figure}[tb]
    \centering
    \includegraphics[width=\linewidth]{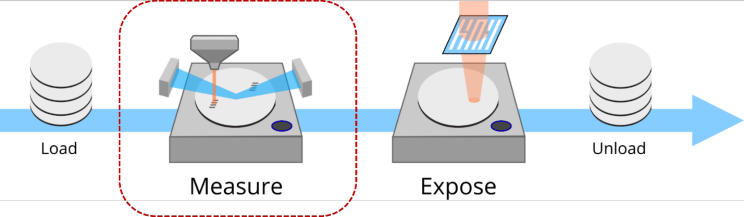}
    \caption{Simpliﬁed view of the life of a wafer in the scanner. The wafer is loaded, measured, exposed, and then unloaded \cite{binns_how_2024}.}
    \label{fig:life_of_a_wafer}
\end{figure}

Our study is conducted at the ASML leveling department. ASML is a global leader in photolithography systems used for semiconductor manufacturing. A critical component of the lithography process is metrology, which involves measuring the wafer with extreme high precision (See Figure~\ref{fig:life_of_a_wafer}). Within the metrology cluster, the leveling department plays an important role by measuring the vertical position of the wafer using level sensors. This precise measurement is utilized to keep the wafer in focus during the exposure stage, thereby ensuring optimal accuracy and precision in the manufacturing process.

Within its leveling department, software plays a critical role, allowing them to process large amounts of data collected by the leveling sensors and perform computations on them. The software in the department is built using the Data Control and Algorithms (DCA) architecture that divides the software into three distinct components: data, control, and algorithms, ensuring that the code base is easily maintainable and has a clear separation of concerns.

\begin{figure}[tb]
    \centering
    \includegraphics[width=\linewidth]{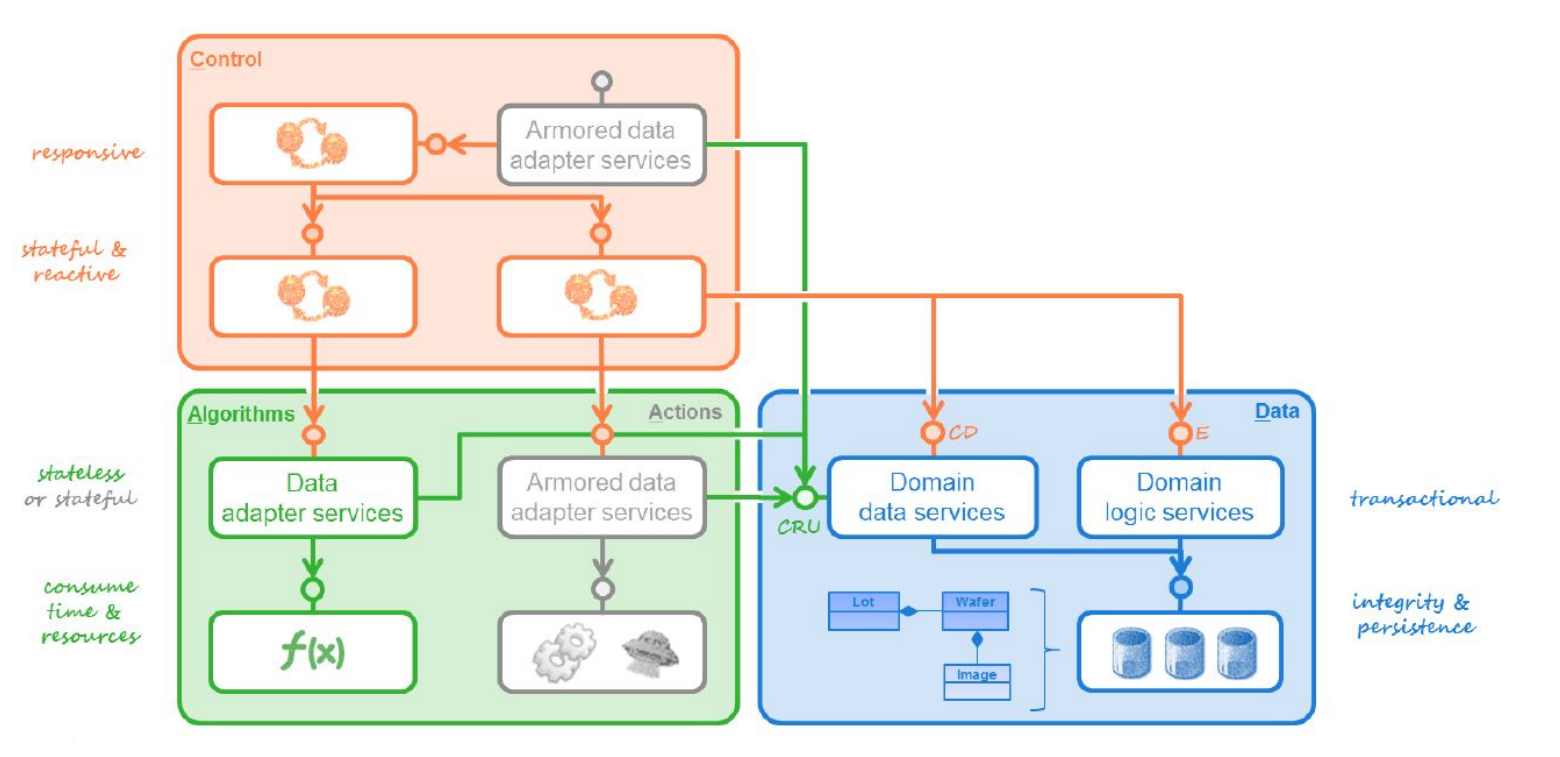}
    \caption{DCA architecture used in the ASML leveling software separating the data, control, and algorithm components \cite{d638b966d1104ff8b3d479a50d07af62}}
    \label{fig:DCA_architecutre}
\end{figure}

The data component is responsible for data persistence, storage, and management. It involves storing and retrieving data related to the wafer and scanner, as well as managing its lifecycle, including creation, update, and deletion. The control component is responsible for decision-making, control logic, and task management. It involves planning, scheduling, and coordinating tasks and subtasks, as well as managing the flow of data and algorithms. It also ensures that the ordering constraints of the system are met. Finally, the algorithm component is responsible for data transformations, calculations, and processing. It involves executing mathematical operations, solving optimization problems, and performing other complex data processing tasks.
An overview of the DCA architecture is presented in Figure~\ref{fig:DCA_architecutre}.

ASML's need for an automated coding tool stems from the repetitive and time-consuming task of writing glue code, which connects various components of the DCA architecture. While this task is relatively straightforward, it can be quite time-consuming for developers, pulling them away from more creative and complex problem-solving activities. Automating these routine coding processes would significantly boost productivity, allowing developers to focus on higher-value work and improving overall efficiency. However, the challenge lies in the specialized terminology and strict interface contracts, which make it difficult for LLMs to effectively handle this type of code.

In this study, we address a key automation challenge: generating new ``\textit{garage}'' interfaces within the data component of the DCA architecture that handle storing and retrieving data from a repository. When the control or algorithm components need specific pieces of data, they access it through these garages. There are three main types: store-garages for storage, retrieve-garages for retrieval, and store-retrieve garages that handle both operations. Although the implementation of a garage follows a fairly standard pattern, it still demands significant manual effort and domain expertise. This makes it an ideal candidate for automation using LLMs, potentially streamlining the process and reducing the workload on developers.

The current coding process presents several challenges that hinder both efficiency and scalability. One issue is the low level of automation: although many garages share similar structures, there is no existing tooling to support their automatic generation based on prior examples. Additionally, code interdependencies further complicate development. Creating a new garage often requires integrating with multiple other components, demanding cross-repository knowledge and careful coordination to avoid breaking builds.

\subsection{Opportunity for Automation}
These challenges make the garage generation task a promising candidate for automation through LLMs. The task exhibits repeated structural patterns that models can learn and reproduce, while its success criteria are clearly defined: the generated garage must compile, pass all tests, and meet ASML’s code quality standards. Moreover, the task reflects broader industrial challenges, where proprietary architectures and domain-specific knowledge play a central role.

\subsection{Research Questions}
To explore the potential of LLMs in automating garage generation, we focus on three key research questions:

\begin{itemize}
    \item \textbf{\textit{RQ1:} To what extent do different prompting techniques affect the performance of generated domain-specific code?}\\
    This question investigates the extent to which various prompting techniques influence the quality and correctness of code generated for proprietary environments. By systematically varying the prompting approach, this research question seeks to identify optimal strategies for generating high-quality and functional code from LLMs. 

    \item \textbf{\textit{RQ2:} How do generic LLMs compare to code-specific LLMs in generating domain-specific, interdependent code?}\\
    This question examines the relative performance of generic LLMs, such as the Gemma models trained on diverse datasets, versus code-specific models like CodeGemma, which are optimized for software engineering tasks. The evaluation focuses on their ability to generate buildable, functionally correct, and high-quality code.

    \item \textbf{\textit{RQ3:} To what extent does the size of the large language model influence the performance of generating domain-specific code?}\\
    Model size is often correlated with improved performance in natural language processing tasks. This question examines whether larger models, with their increased capacity for learning complex patterns, offer tangible benefits for generating domain-specific code. By systematically comparing models of varying sizes, we aim to identify the trade-offs between computational cost and generation quality. Gaining insight into this relationship is crucial for selecting models that balance efficiency and performance in industrial settings.
\end{itemize}

\section{Methodology}
To evaluate the feasibility of using LLMs for domain-specific code generation at ASML, we designed a practical framework for assessing model performance in real industrial settings. The goal of our approach is to generate functional “garage” interfaces in ASML’s leveling codebase using LLMs. To support the generation and evaluation of such garages, we follow a multi-step approach:

\begin{enumerate}
    \item \textbf{Benchmark Dataset Creation}: Extraction of 156 garage examples from ASML’s proprietary codebase, with corresponding context files and test cases.
    \item \textbf{Experimental Setup}: Controlled experiments using multiple prompting strategies, model types, and sizes.
    \item \textbf{Evaluation Strategy}: Quantitative and qualitative assessments using match-based, execution-based, and human-evaluated metrics.
    \item \textbf{Result Analysis}: Statistical aggregation and interpretation of performance outcomes.
\end{enumerate}

\subsection{Benchmark Dataset Creation}
We start off by constructing a benchmark dataset consisting of 156 garages from the leveling repository. Each garage’s file path and implementation were stored in the benchmark. Additionally, unit tests were collected to evaluate functional correctness; however, only 42 garages (approximately 27\%) had associated tests.

To provide contextual information for each garage, we recursively collected all files referenced through import statements, assigning a depth value based on their distance in the import hierarchy. Files directly imported by the garage were given a depth of 1, and so on. To reduce noise, auto-generated files with a depth greater than 2 were excluded. This filtering ensured that only the most relevant human-written files were retained for the benchmark.

Given the limited context window of many LLMs, especially when dealing with long files (often exceeding 2000 lines), we summarized these files using the Qwen2.5-Coder-32B-Instruct model. Additionally, we computed embeddings for all context files using the BGE-M3 model and prioritized them using cosine similarity to the prompt embedding. Finally, token counts were calculated to determine which files could fit within the available context window.

The overall format of the benchmark dataset is shown below:

\begin{minted}[fontsize=\small, breaklines=true, bgcolor=gray!5]{json}
[
  {
    "name": "<Garage_Name>",
    "path": "<Garage_FilePath>",
    "component": "<Garage_Component>",
    "solution_code": "<Garage_Implementation>",
    "related_files": [
      {
        "file_path": "<CF_FilePath>",
        "depth": "<CF_Depth>",
        "implementation": "<CF_Impl>",
        "embedding": "<CF_Embedding>",
        "Qwen2.5-Tokens": "<CF_QwenTokens>",
        "DeepSeek-Tokens": "<CF_DSTokens>",
        "Gemma-Tokens": "<CF_GemmaTokens>"
      }, ..., {}
    ],
    "test_directory": "<Garage_TestDirectory>",
    "test_file_name": "<Garage_TestName>"
  }, ..., {}
]
\end{minted}

\begin{figure}[b]
    \centering
    \includegraphics[width=\linewidth]{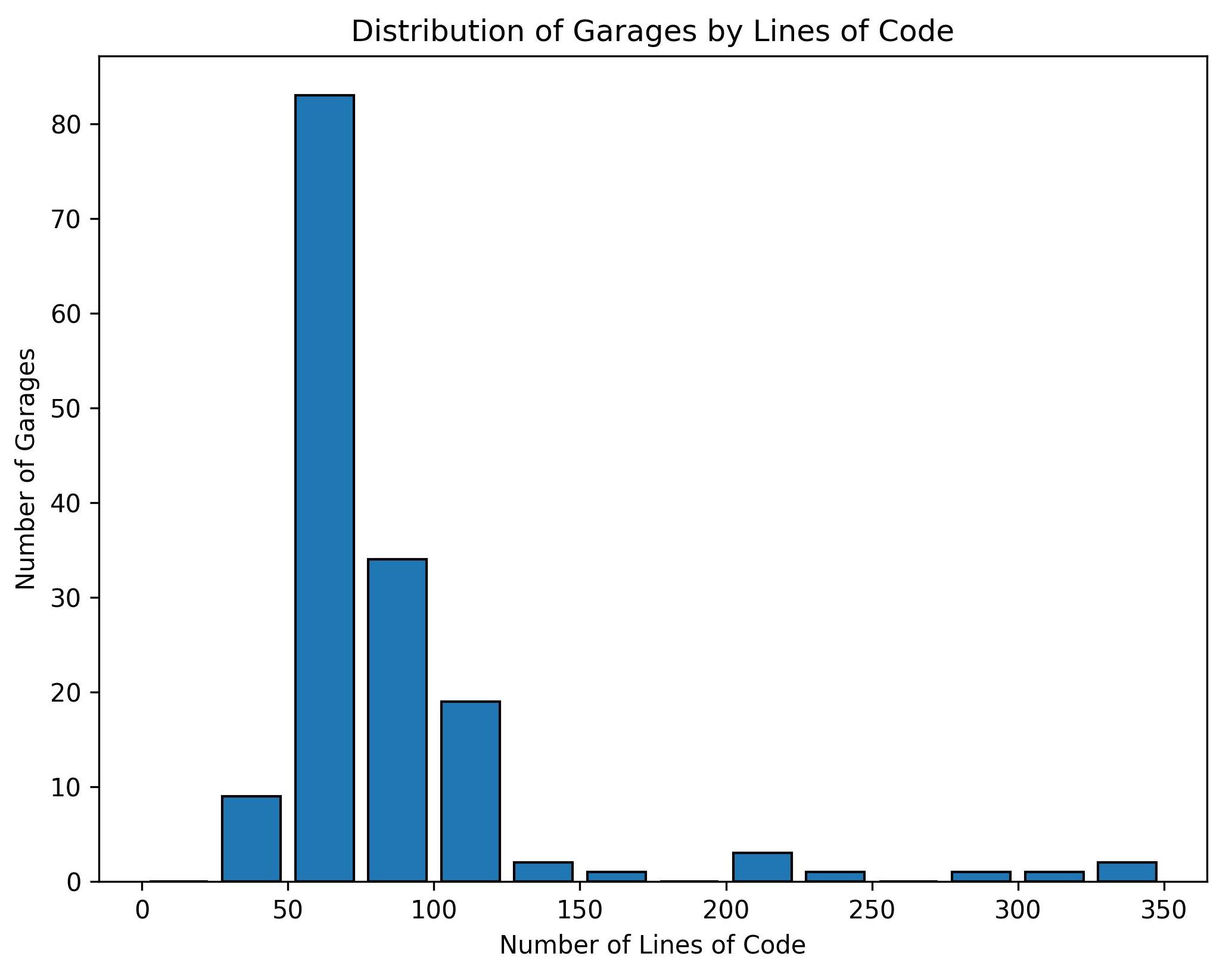}
    \caption{Histogram with the distribution of garages in the benchmark by the number of lines of code.}
    \label{fig:garages/lineofcode}
\end{figure}

The garages we collected for the benchmark dataset exhibit a wide range of sizes in terms of lines of code. Figure~\ref{fig:garages/lineofcode} presents a histogram illustrating the distribution of garage sizes and their corresponding frequencies. As evident from the figure, the majority of garages are small in size, containing less than 100 lines of code. Garages with more than 100 lines of code are less common, with only a small proportion of the benchmark falling into this category.

\subsection{Experimental Setup}
The following section describes the experimental setup that we used to answer the three research questions, providing a clear and concise overview of the study's design and implementation.

\subsubsection{RQ1: Prompting Technique}
The first experiment examined the influence of prompting techniques on generated code quality. Five distinct prompting techniques were selected:
\begin{itemize}
    \item Zero-shot prompting
    \item One-shot prompting
    \item Few-shot prompting
    \item One-shot chain-of-thought prompting
    \item Few-shot chain-of-thought prompting
\end{itemize}

Zero-shot prompting~\cite{kojima_large_2023} involves instructing a language model to perform a task using only a task description, without providing any examples. The model relies entirely on its pre-trained knowledge to generate a response. In contrast, one-shot and few-shot prompting~\cite{brown_language_2020} enhance task understanding by including one or a few input-output examples, respectively. These examples serve as demonstrations to guide the model’s behavior. While few-shot prompting can significantly improve performance on complex tasks, it also increases input length and is sensitive to the quality and selection of examples.

Chain-of-thought (CoT) prompting~\cite{wei_chain--thought_2023} builds on these techniques by encouraging the model to generate intermediate reasoning steps. In one-shot CoT prompting, a single example includes both the task and a step-by-step reasoning process, while few-shot CoT prompting provides multiple such examples. This structured approach helps larger models perform better on tasks requiring logical reasoning, such as arithmetic or commonsense inference. 
% CoT prompting leverages emergent abilities like coherence and symbolic reasoning, making it especially effective for problems that benefit from explicit, multi-step thinking.

All experiments for research question 1 were conducted using the Qwen2.5-Coder-32B-Instruct model with Q8\_0 quantization and a 32k token context window, using default sampling parameters (temperature: 0.7, top-p: 0.8, top-k: 20). To manage the limited context window, we allocated 25k tokens for context files and 7k tokens for the prompt and output, based on the size of the largest prompt and garage in the dataset. Context files were prioritized first by import depth and then by cosine similarity to the prompt embedding, ensuring the most relevant files were included.

\subsubsection{RQ2: Generic versus Code-specific LLMs}
The second experiment compared the performance of generic LLMs and code-specific LLMs in generating garages within ASML. To ensure a fair comparison, we used code-specific models that were fine-tuned from the same generic base models, maintaining identical architectures. The models used for the experiments, as presented below, were evaluated in their default configurations using their standard sampling parameters with Q8\_0 quantization.

\begin{itemize}
    \item Qwen2.5-7B-Instruct (7B)
    \item Qwen2.5-Coder-7B-Instruct (7B)
    \item gemma-7b-it (7B)
    \item codegemma-7b-it (7B)
    \item DeepSeek-V2-Lite-Chat (16B)
    \item DeepSeek-Coder-V2-Lite-Instruct (16B)
\end{itemize}

All experiments used single-shot Chain-of-Thought (CoT) prompting due to its effectiveness in code generation~\cite{yu_towards_2023}. Context file selection followed the same prioritization strategy as in the first experiment, with Qwen2.5 and DeepSeek models utilizing a 32k token context window, while Gemma models were limited to 8k tokens. 

\subsubsection{RQ3: Model Size}
The third and final experiment that we conducted investigated the impact of large language model size variations on code generation performance within the ASML leveling department. To ensure a fair comparison, all models were selected from the same family, guaranteeing that the architecture remained consistent while the primary difference was the model size.

\begin{itemize}
    \item Qwen2.5-Coder-0.5B-instruct
    \item Qwen2.5-Coder-1.5B-instruct
    \item Qwen2.5-Coder-3B-instruct
    \item Qwen2.5-Coder-7B-instruct
    \item Qwen2.5-Coder-14B-instruct
    \item Qwen2.5-Coder-32B-instruct 
\end{itemize}

All models were configured uniformly using Q8\_0 quantization, a 32k token context window, and default sampling parameters (temperature: 0.7, top-p: 0.8, top-k: 20), with 25k tokens allocated for context files and 7k for the prompt and output. Context file prioritization followed the same methodology as in the first two experiments, ensuring consistency across setups. Single-shot chain-of-thought prompting was used throughout, based on its demonstrated effectiveness in prior research~\cite{yu_towards_2023}. 

\subsection{Evaluation Strategy}
To assess the quality of generated code, we use a three-phase evaluation approach.

\begin{itemize}
    \item \textbf{Match-based Metrics:} Assess the similarity between generated code and a reference implementation by comparing tokens, syntax structures, or exact outputs.
    \item \textbf{Execution-based Metrics:} Evaluate the functional correctness of the generated code by compiling and executing it, then comparing the observed behavior and outputs against expected results.
    \item \textbf{Manual Evaluation:} Involve human judgment to assess qualitative aspects such as readability, maintainability, and adherence to coding standards—factors not fully captured by automated similarity or execution-based metrics.
\end{itemize}

\subsubsection{Match-based Metrics}
The match-based metrics used in this study to compare the generated output and the reference implementation are BLEU, CodeBLEU, and ROUGE. The BLEU score~\cite{papineni_bleu_2002} is calculated based on the number of matching n-grams between the generated code and the reference code. It then calculates the weighted mean of these matches to get the overall similarity score. CodeBLEU~\cite{ren_codebleu_2020} enhances the traditional BLEU score by incorporating syntactic and semantic information specific to code. This is achieved by leveraging abstract syntax trees to represent code syntax and data-flow to capture code semantics. Finally, the ROUGE~\cite{lin_rouge_2004} score takes the recall value of n-gram overlaps between the generated code and the reference code. It provides a measure of how well the generated code matches the reference code in terms of content and structure.

\subsubsection{Execution-based Metrics}
Since code-similarity metrics offer limited insight into the functional correctness of the generated code, the second stage of the evaluation process involved assessing execution-based performance metrics. We use three main execution based metrics:

\begin{itemize}
    \item \textbf{Buildability:} While traditional execution-based metrics, as noted by Chen et al. \cite{chen_survey_2025}, typically focus on file- or function-level evaluations, but often neglect repository-level analysis. Consequently, evaluating the buildability of the generated code was deemed essential. To the best of our knowledge, there is no existing metric that assesses the buildability of generated code; therefore, we propose a novel performance metric, termed \textit{build@k}, which draws inspiration from the pass@k metric \cite{kulal_spoc_2019}. The \textit{build@k} metric takes $k$ code samples per garage. A garage is considered buildable if it successfully integrates into the project and passes the build pipelines; the total fraction of buildable garages is reported. Formula~\ref{formula:buildk} as given below formally defines the \textit{build@k} metric. 

    \begin{equation} \label{formula:buildk} 
            \text{build@k} = \frac{1}{N} \sum_{i=1}^{N} \begin{cases}
            1 & \begin{aligned}
              & \text{if any of the $k$ generated code} \\
              & \text{instances for garage$_i$ is built} \\
              \end{aligned} \\
            0 & \text{otherwise}
            \end{cases}
    \end{equation}

    \item \textbf{Unit Tests:} To assess the functional correctness of the generated code, we execute the unit tests associated with the garage and employ the \textit{pass@k} metric, as proposed by Kulal et al.~\cite{kulal_spoc_2019}. The \textit{pass@k} metric reports the total fraction of garages that are successfully solved (only consider garages with unit tests are considered for this metric). Formula \ref{formula:passk} as given below formally defines the \textit{pass@k} metric.

    \begin{equation}\label{formula:passk}
        \text{pass@k} = \frac{1}{N} \sum_{i=1}^{N} \begin{cases}
            1 & \begin{aligned}
              & \text{if any of the $k$ generated } \\
              & \text{code instances for garage$_i$} \\
              & \text{passes all unit tests} \\
              \end{aligned} \\
            0 & \text{otherwise}
            \end{cases}
    \end{equation}

    \item \textbf{Code Quality:} A significant gap in existing research is the evaluation of generated code in terms of code quality aspects, such as maintainability, readability, and performance efficiency~\cite{chen_survey_2025}. We use the TICS tool, an evaluation platform used at ASML~\cite{jansen_tomtom_nodate}, to assess the quality of the generated code. The TICS tool (created by TIOBE) provides a comprehensive evaluation of code quality, enabling us to gauge the maintainability and readability of the generated code. This tool provides a more detailed assessment of the generated code's quality, moving beyond just functional correctness. 
    
    The TICS tool assesses the generated code across a wide range of categories, including class interface, code organization, error handling, naming, and many others. Each rule is assigned a severity level, ranging from 1 to 10, with level 1 indicating the most critical issues (program errors) and level 10 representing the least critical rules (style issues). Issues assigned to levels 1-7 are deemed critical, whereas those assigned to levels 8-10 are considered non-critical \cite{jansen_coding_nodate}. Using the TICS score, we will measure two key metrics: the average number of violations per buildable garage and the average number of critical violations (levels 1–7) per buildable garage. 
\end{itemize}

\subsubsection{Manual Evaluation}
As the final step in our evaluation strategy, manual evaluation enables us to assess aspects of generated code such as readability, maintainability, and adherence to coding standards and best practices, which are not fully captured by code-similarity and execution-based metrics. 

\section{Results}
\subsection{RQ1: Prompting Techniques}
\subsubsection{Match-based Metrics}
The CodeBLEU similarity results are presented in Table~\ref{table:bleu_codebleu_experiment1}, highlighting clear differences in performance across prompting techniques. Zero-shot and one-shot prompting yielded the lowest similarity scores across BLEU, CodeBLEU, and ROUGE metrics, indicating limited effectiveness in generating domain-specific code. In contrast, few-shot, one-shot CoT, and few-shot CoT prompting performed significantly better, with few-shot CoT leading. BLEU and ROUGE scores are not reported due to space limitations; however, they exhibited similar trends to the CodeBLEU scores.

Across all prompting methods, similarity scores improved as the number of generated outputs (\(k\)) increased, particularly from \(k=1\) to \(k=3\), suggesting that generating multiple outputs increases the likelihood of producing code closer to the reference. However, the gains between \(k=3\) and \(k=5\) were smaller, indicating diminishing returns with additional generations. 

\begin{table}
    \caption{Mean BLEU and CodeBLEU scores of generated code across five different prompting techniques, with $k$ values of 1, 3, \& 5.}
    \centering
    \begin{tabular}{llll}
                          & \multicolumn{3}{c}{\textbf{CodeBLEU}}                                                                  \\
                          & \multicolumn{1}{c}{\textit{k=1}} & \multicolumn{1}{c}{\textit{k=3}} & \multicolumn{1}{c}{\textit{k=5}} \\ 
    \midrule
    \textbf{Zero-Shot}    & 0.320                            & 0.360                            & 0.380                            \\
    \textbf{One-Shot}     & 0.395                            & 0.440                            & 0.456                            \\
    \textbf{Few-Shot}     & 0.470                            & 0.513                            & 0.531                            \\
    \textbf{One-Shot CoT} & 0.456                            & 0.502                            & 0.512                            \\
    \textbf{Few-Shot CoT} & \textbf{0.483}                   & \textbf{0.523}                   & \textbf{0.534}                  \\
    \bottomrule
    \end{tabular} 
    \label{table:bleu_codebleu_experiment1}
\end{table}

\subsubsection{Execution-based metrics}
In terms of the build@k metric, the results (as presented in table \ref{table:buildk_experiment1}) reveal a clear difference in the ability of the generated code to successfully integrate and compile within the project. Zero-shot prompting consistently performed the worst across all values of \(k\), while few-shot prompting achieved the highest buildability at \(k=1\) and \(k=5\), generating 44 buildable garages at the highest setting. One-shot CoT prompting showed strong performance, particularly at \(k=3\), where it outperformed all other techniques. Overall, build@k scores improved with higher \(k\) values across all prompting methods, indicating that generating multiple outputs increases the likelihood of producing buildable code. However, the rate of improvement varied, with the CoT-based prompting techniques showing smaller gains between \(k=3\) and \(k=5\).

\begin{table}[tb]
    \caption{build@k score of generated code across five different prompting techniques with $k$ values of 1, 3, and 5.}
    \centering
    \begin{tabular}{llll}
    & \multicolumn{3}{c}{\textbf{build@k}} \\ \midrule
    & \multicolumn{1}{c}{\textit{k=1}} & \multicolumn{1}{c}{\textit{k=3}} & \multicolumn{1}{c}{\textit{k=5}} \\ \midrule
    \textbf{Zero-Shot Prompting}    & 0.01935                          & 0.04516                          & 0.08333                          \\
    \textbf{One-Shot Prompting}     & 0.07189                          & 0.14379                          & 0.24837                          \\
    \textbf{Few-Shot Prompting}     & \textbf{0.12820}                 & 0.21795                          & \textbf{0.28205}                 \\
    \textbf{One-Shot CoT Prompting} & 0.12179                          & \textbf{0.23077}                 & 0.23718                          \\
    \textbf{Few-Shot CoT Prompting} & 0.10897                          & 0.20513                          & 0.21154    
    \\ \bottomrule
    \end{tabular}
    \label{table:buildk_experiment1}
\end{table}

The unit test results from Experiment 1 revealed that none of the generated garages passed the available tests, resulting in a pass@k score of 0.0 across all prompting techniques and values of \(k\). This outcome was primarily due to two factors: a lack in the model's ability to generate garages that successfully build within the project, and a lack of unit test coverage among those that did. In most cases, the majority of buildable garages lacked associated unit tests, with coverage ranging from 0\% to 21\% depending on the prompting technique and \(k\) value.

Finally, in terms of code quality, the results presented in Table~\ref{table:codequality_k5_experiment1} show that zero-shot prompting consistently produced the lowest-quality code across all \(k\) values. It had the highest violations per build, ranging from 2.00 to 2.29, and also introduced critical violations at \(k=3\) (0.143 per build), while all other techniques maintained lower rates. In contrast, one-shot CoT and few-shot CoT prompting achieved the best results, with violations per build close to or exactly 1.00 at \(k=1\), and minimal critical violations across all \(k\) values. Overall, the results indicate that prompts containing examples improve the quality of the generated code. 

\begin{table}[tb]
\caption{Code quality metrics of generated code across five different prompting techniques. \textbf{VPB} stands for violations per build.}
\centering
\begin{tabular}{ll|lllll}
    &                                        & \rotatebox{90}{\textbf{Zero-Shot}}           & \rotatebox{90}{\textbf{One-Shot}}                    & \rotatebox{90}{\textbf{Few-Shot}} & \rotatebox{90}{\textbf{One-Shot CoT}}                 & \rotatebox{90}{\textbf{Few-Shot CoT}}                \\
\midrule
    & \textbf{VPB}          & {\color[HTML]{FE0000} 2.00}  & 1.27                                 & 1.05              & {\color[HTML]{32CB00} \textbf{1.00}}  & {\color[HTML]{32CB00} \textbf{1.00}} \\
\multirow{-2}{*}{\textit{\textbf{k=1}}} & \textbf{Critical VPB} & 0.00                        & 0.00                                & 0.00             & 0.00                                 & 0.00                                \\
\midrule
    & \textbf{VPB}          & {\color[HTML]{FE0000} 2.29}  & 1.32                                 & 1.15              & {\color[HTML]{32CB00} \textbf{1.06}}  & 1.13                                 \\
\multirow{-2}{*}{\textit{\textbf{k=3}}} & \textbf{Critical VPB} & {\color[HTML]{FE0000} 0.143} & 0.05                                & 0.03             & {\color[HTML]{32CB00} \textbf{0.03}} & 0.06                                \\
\midrule
    & \textbf{VPB}          & {\color[HTML]{FE0000} 2.00}  & {\color[HTML]{32CB00} \textbf{1.11}} & 1.14              & {\color[HTML]{32CB00} \textbf{1.11}}  & 1.12                                 \\
\multirow{-2}{*}{\textit{\textbf{k=5}}} & \textbf{Critical VPB} & 0.00                        & 0.00                                & 0.05             & 0.05                                 & {\color[HTML]{FE0000} 0.06}        \\
\bottomrule
\end{tabular}
\label{table:codequality_k5_experiment1}
\end{table}

\subsection{RQ2: Code-Specific versus Generic LLMs}

\subsubsection{Match-based metrics}
\begin{figure}[tb]
    \centering
    \includegraphics[width=\linewidth]{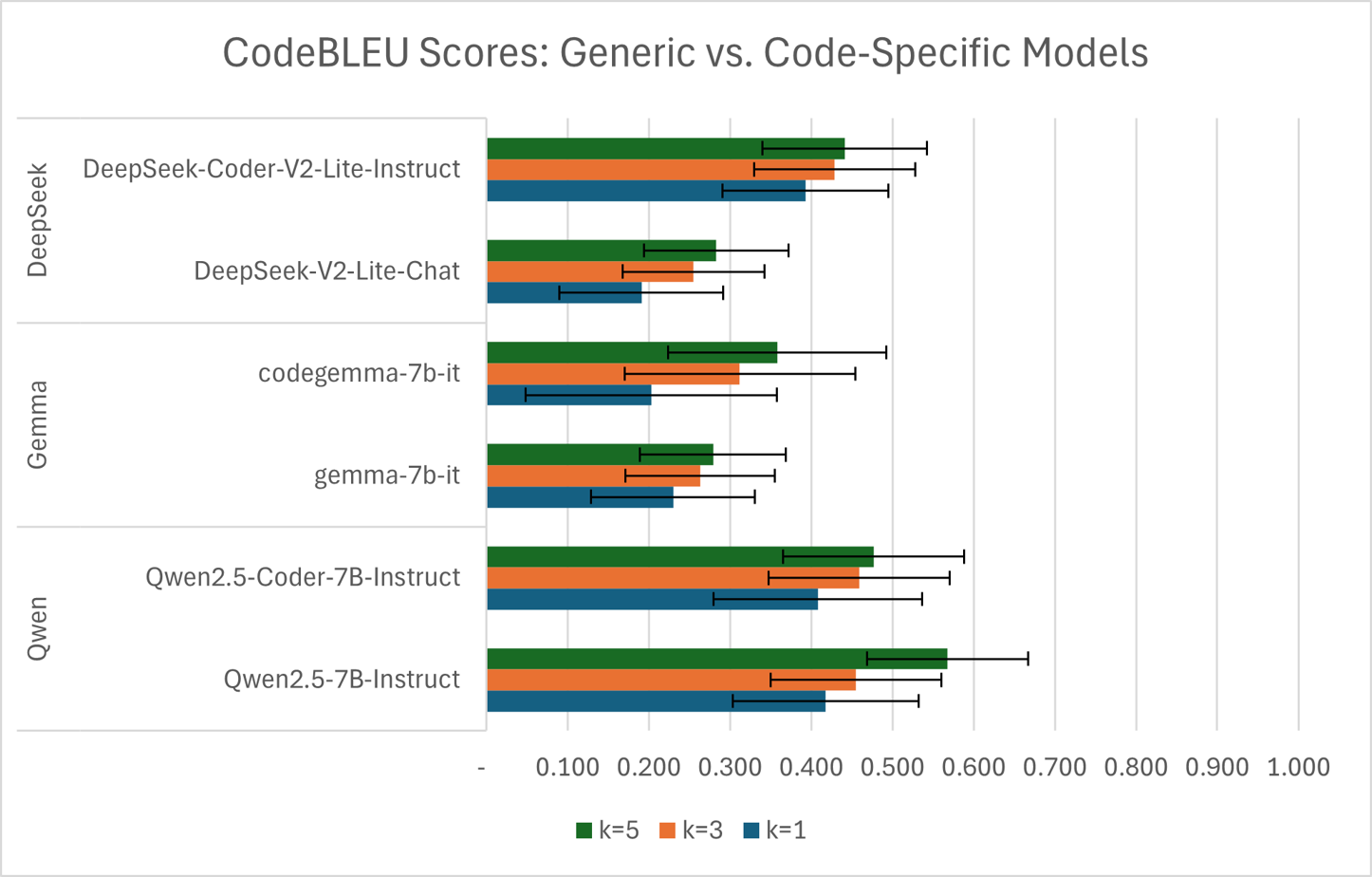}
    \caption{CodeBLEU score of generated code using code-specific and generic LLMs, with k values of 1, 3, and 5}
    \label{fig:experiment2_codebleu}
\end{figure}

The CodeBLEU scores (see figure \ref{fig:experiment2_codebleu}) demonstrate that overall code-specific LLMs outperform their generic counterparts across all three model families; Qwen2.5, Gemma, and DeepSeek-V2. These code-specific models generate code that more closely matches the reference implementation. While the Qwen2.5 models showed minimal differences between the generic and code-specific variants, the Gemma and DeepSeek-V2 families exhibited substantial improvements with the code-specific fine-tuning. Notably, the DeepSeek-Coder-V2-Lite-Instruct model achieved a 105.8\% increase in CodeBLEU score over its generic counterpart at \(k=1\). BLEU and ROUGE scores exhibited similar trends and have therefore been omitted for brevity.

\subsubsection{Execution-based metrics}
\begin{figure}[tb]
    \centering
    \includegraphics[width=\linewidth]{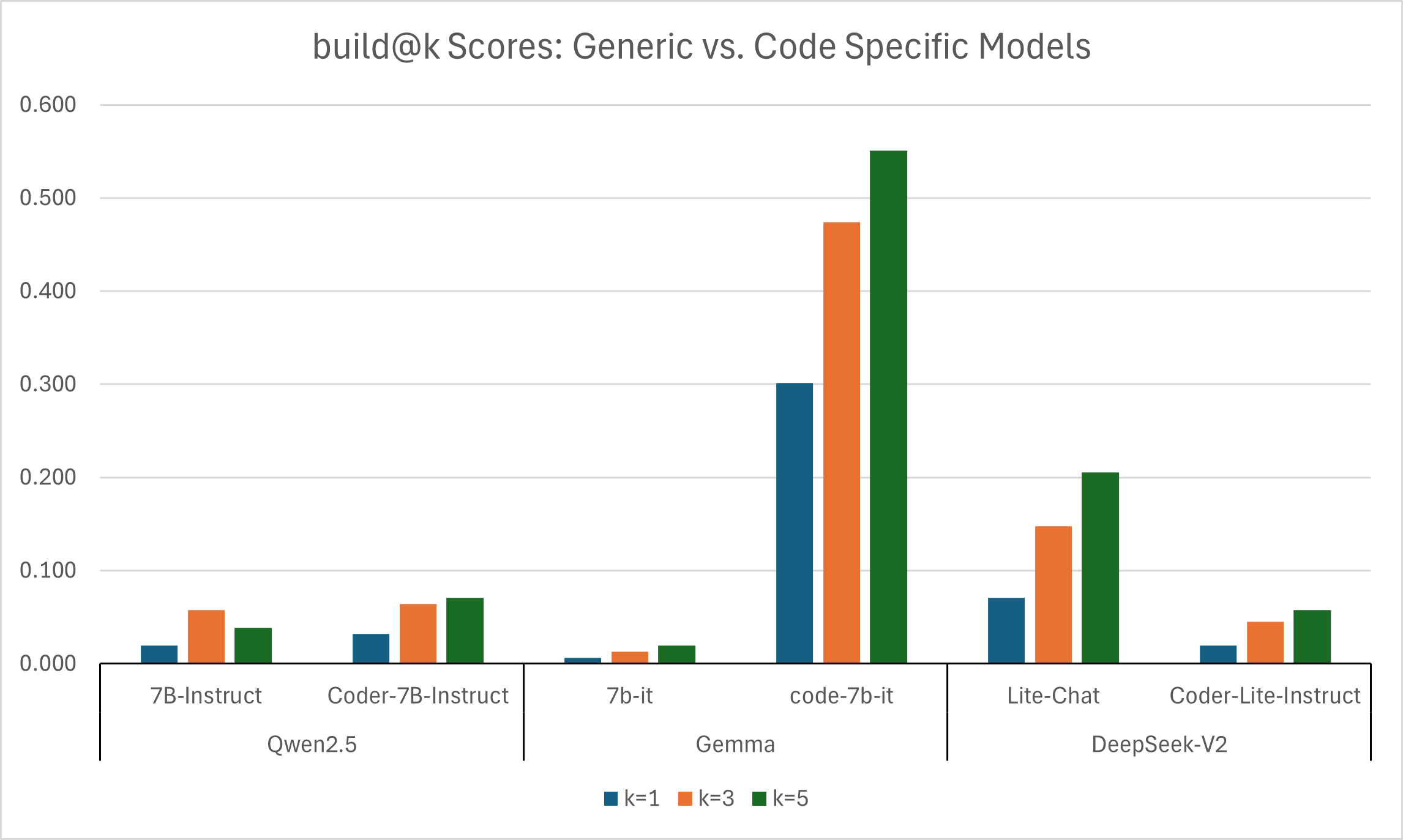}
    \caption{Bar chart with build@k score of generated code comparing generic and code-specific LLMs for the Qwen2.5, Gemma, and DeepSeek-V2 families, with $k$ values of 1, 3, and 5.}
    \label{fig:buildk_experiment2}
\end{figure}

In terms of the build@k scores, the code-specific Qwen2.5 model slightly outperformed its generic counterpart across all \(k\) values, though the differences were marginal. In contrast, the Gemma models exhibited a substantial gap in buildability: the code-specific variant generated significantly more buildable garages than the generic model, with a build@k score of 0.551 at \(k=5\) compared to just 0.019. Interestingly, the DeepSeek-V2 models showed the opposite pattern, where the generic model consistently achieved higher build@k scores than its code-specific counterpart. Despite the code-specific DeepSeek model producing more similar code, its lower buildability suggests that similarity alone is not a sufficient indicator of functional correctness, and that model suitability may vary depending on the evaluation criteria. Figure \ref{fig:buildk_experiment2} presents the build@k scores from this experiment in a bar chart, providing a visual overview of the build performance of the generated code. 

In terms of functional correctness, none of the generated garages successfully passed the unit tests, resulting in a score of 0 across all models and \(k\) values. Unit test coverage was generally low across all model families, with only 28\% of buildable garages from the Qwen2.5 models and 26\% from the DeepSeek-V2 models having associated tests. The Gemma models showed the highest coverage at approximately 50\%, yet still failed to produce any passing outputs. These results highlight a persistent gap in the functional correctness of generated code.

The code quality results from the second experiment reveal varying trends across the three model families. For the Qwen2.5 models, the code-specific variant consistently outperformed the generic model, maintaining a violations per build ratio close to 1.00 and incurring no critical violations across all \(k\) values. In contrast, the Gemma models showed the opposite trend: the code-specific model exhibited significantly higher violations per build (exceeding 3.20) and a critical violations per build ratio above 0.55, while the generic model maintained lower values in both metrics. The DeepSeek-V2 models followed a similar pattern to Qwen2.5, with the code-specific model achieving lower violations per build (ranging from 1.43 to 2.00) and no critical violations, compared to the generic model, which consistently exceeded 3.00 violations per build and had a critical violation rate of approximately 0.5. These results suggest that while code-specific fine-tuning can improve code quality, its effectiveness varies depending on the model family.

\begin{table*}[tb]
\caption{Code quality metrics of generated code, comparing code-specific and generic LLMs with $k$ values of 1, 3, and 5. \textbf{VPB} stands for violations per build.}
\centering
\begin{tabular}{ll|ll|ll|ll}
   & & \multicolumn{2}{c|}{\textit{\textbf{k=1}}}                                  & \multicolumn{2}{c|}{\textit{\textbf{k=3}}}                                  & \multicolumn{2}{c}{\textit{\textbf{k=5}}} \\
   & & \multicolumn{1}{c}{\textbf{Generic}} & \multicolumn{1}{c|}{\textbf{Code}}   & \multicolumn{1}{c}{\textbf{Generic}} & \multicolumn{1}{c|}{\textbf{Code}}   & \multicolumn{1}{c}{\textbf{Generic}} & \multicolumn{1}{c}{\textbf{Code}}    \\
\midrule
   & \textbf{VPB}          & 1.67                                 & {\color[HTML]{32CB00} \textbf{1.00}} & 1.67                                 & {\color[HTML]{32CB00} \textbf{1.00}} & 1.67                                 & {\color[HTML]{32CB00} \textbf{1.09}} \\
\multirow{-2}{*}{\textit{Qwen2.5}}     & \textbf{Critical VPB} & 0.00                                & 0.00                                & 0.00                                & 0.00                                & 0.00                                & 0.00                                \\
\midrule
   & \textbf{VPB}          & {\color[HTML]{32CB00} \textbf{2.00}} & 3.60                                 & {\color[HTML]{32CB00} \textbf{2.00}} & 3.43                                 & {\color[HTML]{32CB00} \textbf{2.00}} & 3.20                                 \\
\multirow{-2}{*}{\textit{Gemma}}       & \textbf{Critical VPB} & 0.00                                & 0.72                                & 0.00                                & 0.65                                & 0.00                                & 0.56                                \\
\midrule
   & \textbf{VPB}          & 3.27                                 & {\color[HTML]{32CB00} \textbf{2.00}} & 3.22                                 & {\color[HTML]{32CB00} \textbf{1.43}} & 3.03                                 & {\color[HTML]{32CB00} \textbf{1.56}} \\
\multirow{-2}{*}{\textit{DeepSeek-V2}} & \textbf{Critical VPB} & 0.55                                & 0.00                                & 0.57                                & 0.00                                & 0.47                                & 0.00                                \\
\bottomrule
\end{tabular}
\label{table:experiment2_code_quality}
\end{table*}

\subsection{RQ3: Model Size}
\subsubsection{Match-based metrics}
\begin{figure}[tb]
    \centering
    \includegraphics[width=\linewidth]{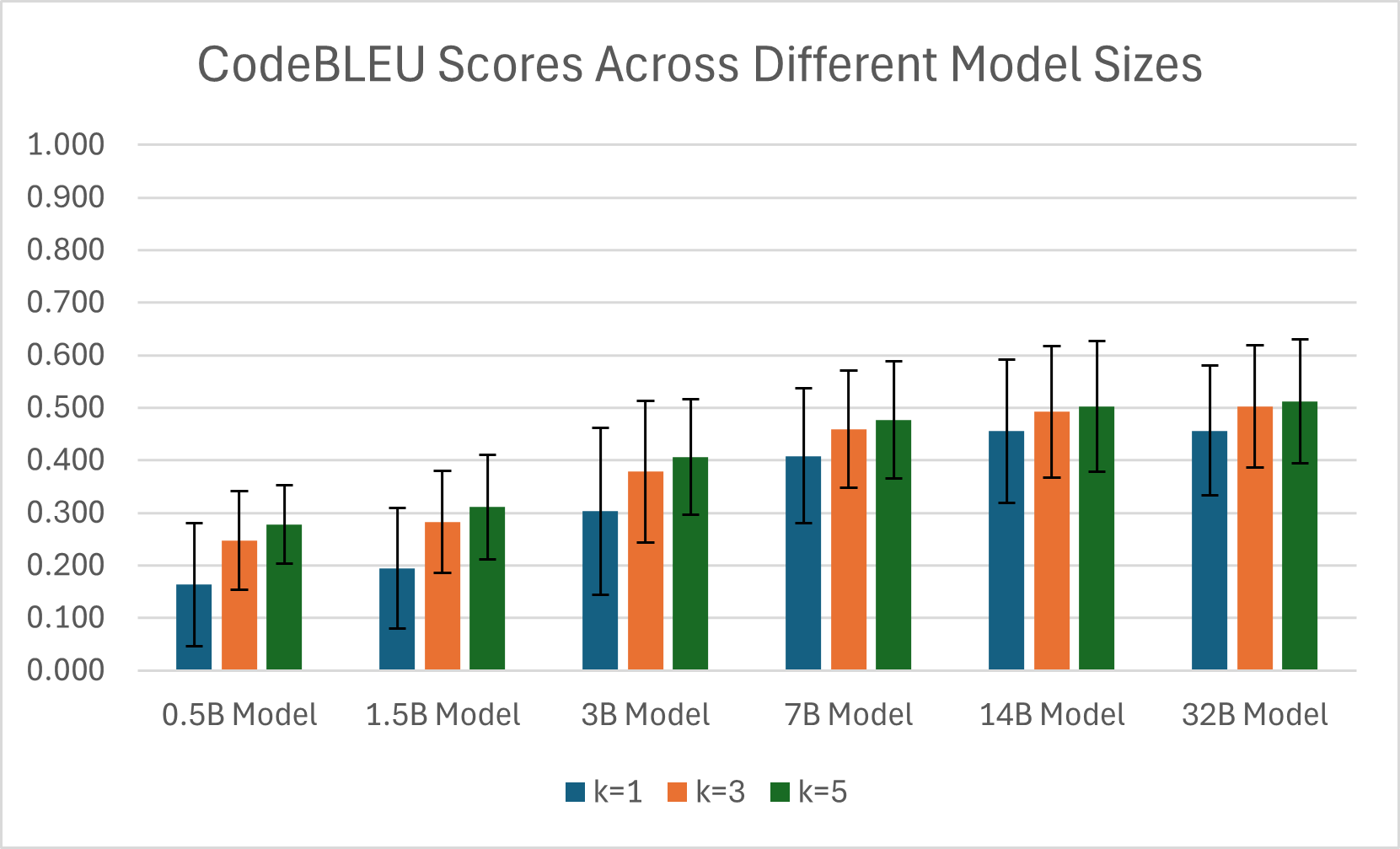}
    \caption{Bar chart with mean CodeBLEU scores of generated code across different model sizes of Qwen2.5-Coder-Instruct, with $k$ values of 1, 3, \& 5.}
    \label{fig:codebleu_experiment3}
\end{figure}

The CodeBLEU scores from experiment 3, as shown in Figure~\ref{fig:codebleu_experiment3}, indicate a positive correlation between model size and similarity. As the number of parameters increases, the generated code exhibits higher similarity to the reference solutions. The most significant improvements occur between the 0.5B and 3B models, with the rate of improvement reducing beyond the 14B model. The 32B model achieves the highest CodeBLEU scores across all $k$ values, although the performance gain over the 14B model is marginal, suggesting a plateauing effect at larger scales. Additionally, increasing the $k$ value from 1 to 3 yields noticeable improvements in similarity, while the difference between $k=3$ and $k=5$ is less substantial. These trends are consistent with those observed for BLEU and ROUGE, reinforcing the conclusion that larger models and moderate increases in $k$ enhance code generation quality, albeit with diminishing returns at the upper end of the model size spectrum.

\subsubsection{Execution-based metrics}
\begin{figure}[tb]
    \centering
    \includegraphics[width=\linewidth]{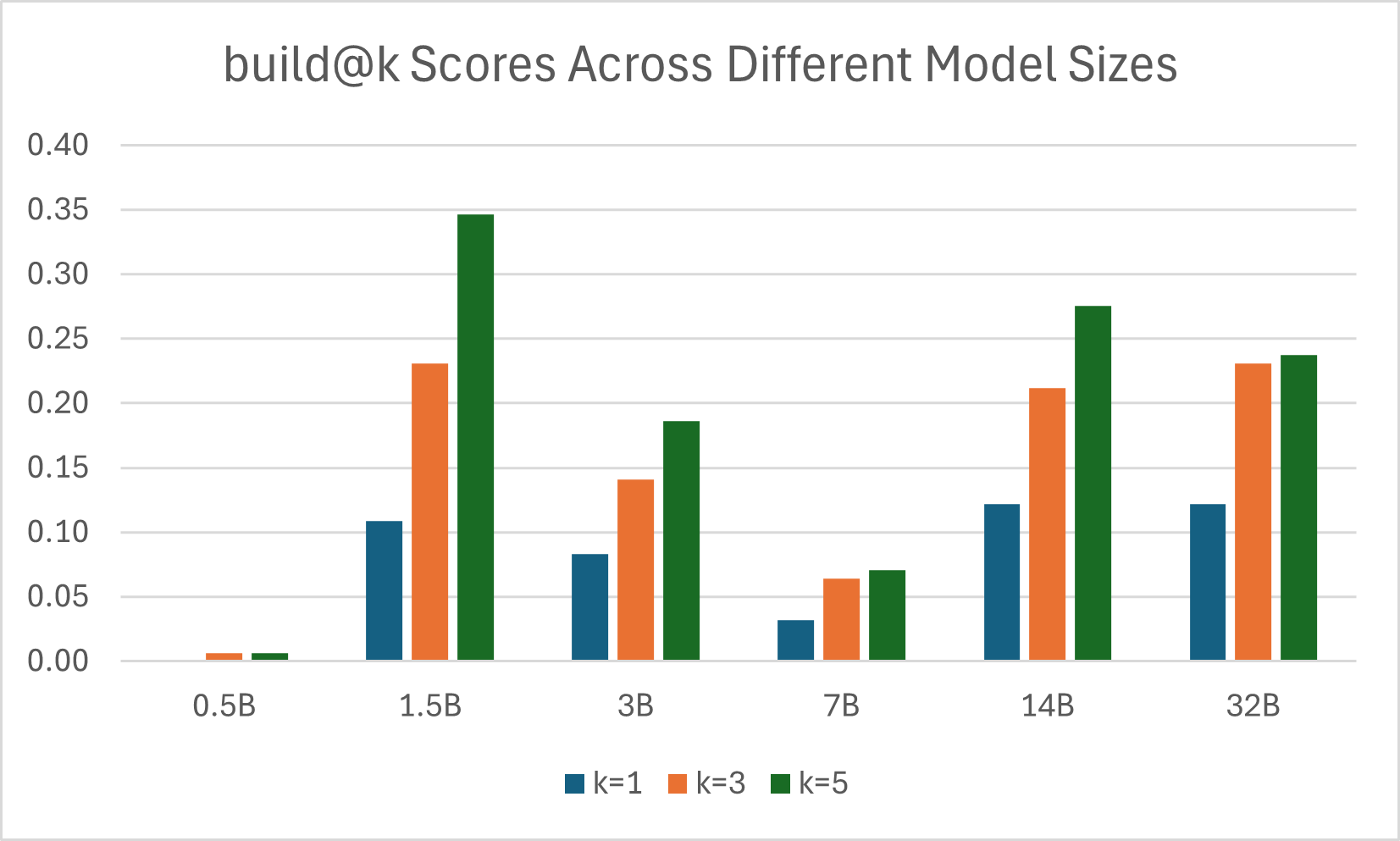}
    \caption{Bar chart with build@k score of generated code across different model sizes of Qwen2.5-Coder-Instruct, with $k$ values of 1, 3, \& 5.}
    \label{fig:buildk_experiment3}
\end{figure}

Figure~\ref{fig:buildk_experiment3} reveals that the build@k scores vary across different model sizes, indicating that buildability is not strictly correlated with model scale. Unlike the match-based metrics, which showed a consistent upward trend with increasing model size, the build@k scores fluctuate, with some smaller models outperforming larger ones. Notably, the 1.5B model achieved the highest build@k scores at $k=3$ and $k=5$, even surpassing the 14B and 32B models. The 0.5B model performed the worst, failing to produce any buildable code at $k=1$ and only marginally improving at higher $k$ values. While the 14B and 32B models showed strong performance at $k=1$,  the 1.5B model managed to generate more buildable garages at higher $k$ values. Overall, increasing $k$ generally improved buildability across all models except the 0.5B variant, likely due to the increased chances of generating at least one buildable sample. 

Despite some garages being successfully built, none of the generated outputs passed the unit tests, indicating a lack of functional correctness. Since only 27\% of the built garages had associated unit tests, it was not possible to draw firm conclusions about the models' ability to generate functionally correct code.

The final evaluation assesses the code quality of the generated outputs (Table~\ref{table:experiment3_code_quality}). The results show a clear distinction between smaller and larger models. The 0.5B, 1.5B, and 3B models consistently produce code with higher violation rates, exceeding a ratio of 2 violations per build across all $k$ values, and also exhibit elevated critical violation ratios, up to 1.0 in the case of the 0.5B model at $k=5$. In contrast, the larger models (7B, 14B, and 32B) maintain a violations per build ratio close to 1.0 and demonstrate minimal critical violations, often registering 0. These findings suggest that larger models not only generate more syntactically correct code but also adhere more closely to coding standards and best practices, resulting in cleaner and safer outputs.

\begin{table}[tb]
    \caption{Code quality metrics of generated code across different model sizes of Qwen2.5-Coder-Instruct, at k = 1, 3, and 5. \textbf{VPB} stands for violations per build.}
    \centering  
    \begin{tabular}{ll|cccccc}
    % \toprule
       & & \multicolumn{6}{c}{\textit{\textbf{Qwen2.5-Coder-Instruct}}} \\ \midrule
       &                                        & \textbf{0.5B}               & \textbf{1.5B}               & \textbf{3B}                 & \textbf{7B}                          & \textbf{14B} & \textbf{32B}                         \\ \midrule
       & \textbf{VPB}          & N/A                         & {\color[HTML]{FE0000} 3.06} & 2.31                        & {\color[HTML]{32CB00} \textbf{1.00}} & 1.05         & {\color[HTML]{32CB00} \textbf{1.00}} \\
\multirow{-2}{*}{\textit{k=1}} & \textbf{Critical VPB} & N/A                         & 0.47                       & 0.15                       & 0.00                                & 0.00        & 0.00                                \\ \midrule
       & \textbf{VPB}          & 2.00                        & 2.81                        & {\color[HTML]{FE0000} 3.68} & {\color[HTML]{32CB00} \textbf{1.00}} & 1.09         & 1.06                                 \\
\multirow{-2}{*}{\textit{k=3}} & \textbf{Critical VPB} & 0.00                       & 0.42                       & 0.91                       & 0.00                                & 0.00        & 0.03                                \\ \midrule
       & \textbf{VPB}          & {\color[HTML]{FE0000} 4.00} & 3.37                        & 2.62                        & {\color[HTML]{32CB00} \textbf{1.09}} & 1.12         & 1.11                                 \\
\multirow{-2}{*}{\textit{k=5}} & \textbf{Critical VPB} & 1.00                       & 0.61                       & 0.31                       & 0.00                                & 0.02        & 0.05 \\ \bottomrule
    \end{tabular}
    \label{table:experiment3_code_quality}
\end{table}

\section{Discussion}
The findings from the three experiments provide valuable insights into the effectiveness of prompting strategies, the comparative performance of generic versus code-specific LLMs, and the influence of model size on code generation quality.

\subsection{Prompting Strategies}
The results from Experiment 1 demonstrate that prompting strategies significantly influence the quality and buildability of generated code. Overall, it was observed that prompts with examples perform substantially better than prompts without examples. Zero-shot prompting consistently underperformed across all metrics, highlighting the limitations of relying solely on task descriptions without contextual examples. In contrast, few-shot and one-shot chain-of-thought (CoT) prompting techniques yielded superior results, particularly in match-based metrics such as CodeBLEU. These techniques leverage the in-context learning ability of the model, enabling it to better infer task structure and domain-specific patterns.

Interestingly, while CoT prompting was expected to enhance reasoning and structural coherence, its performance gains were marginal compared to standard few-shot prompting. In some cases, excessive reasoning chains may have introduced noise or led to overfitting on the examples, reducing generalization. This suggests that the benefits of CoT prompting may be task-dependent and that a single, well-crafted example may be more effective than multiple reasoning steps. 

Upon manual inspection of the generated code, we observed that when prompts lack concrete examples of the desired coding style, the models tend to fall back on common patterns observed in their training data. For instance, generated methods often returned \texttt{nullptr} instead of raising exceptions or using optional objects, which contradicts the expected practice at ASML. Similarly, vectors were frequently constructed without pre-allocating memory, potentially impacting performance. These behaviors were especially prominent with zero-shot prompting, where the absence of guiding examples led to less aligned implementations. 
While the test coverage of the generated garages was low, manual review indicated that many of the garages without unit tests were functionally correct. This suggests that the reported metrics may underestimate the models’ effectiveness in generating valid code.

\subsection{Generic versus Code-Specific LLMs}
Experiment 2 revealed that code-specific LLMs generally outperform their generic counterparts, though the extent of this advantage varies across model families. For the Qwen2.5 models, the performance gap was minimal, indicating that the generic model already possessed strong code generation capabilities. However, in the Gemma and DeepSeek families, the code-specific models demonstrated improvements in both similarity and buildability metrics.

Manual inspection of the generated code revealed several notable limitations across the models. The generic Gemma model frequently reproduced the example garage from the prompt verbatim, suggesting limited generative capability. Both the code-specific Gemma model and the generic DeepSeek-V2 model often produced only pure virtual functions without any functional implementation, reflecting a lack of generalization and a shallow understanding of the task. This behavior also contributed to inflated build@k scores, as the code was syntactically valid but lacked meaningful functionality. In some cases, the generic DeepSeek-V2 model declined to generate code altogether, citing the task's complexity. Overall, the code-specific models showed a better grasp of the prompt, but still struggled to produce usable, fully implemented code.

\subsection{Impact of Model Size}
Our findings confirmed that model size plays a critical role in the performance of LLMs for domain-specific code generation. Larger models consistently achieved higher similarity scores and generated more buildable code. The 14B and 32B models performed best overall, though the marginal gains between them suggest a reduced rate of improvement beyond a certain parameter threshold.

Interestingly, the 1.5B model achieved a high build@k score but often generated non-functional code consisting of virtual methods. This highlights a key limitation of using buildability as a proxy for correctness. Larger models, particularly those above 7B parameters, were more capable of generating functional code. Many garages generated by the 0.5B model lacked actual functionality and instead consisted of long lists of imports, a phenomenon known as the ``repeat curse''~\cite{yao_understanding_2025}, which is a common limitation of smaller-sized LLMs.

\subsection{Threats to Validity}
Several threats to validity were identified across all experiments and can be grouped into three main categories: internal, external, and construct validity threats.

\subsubsection{Internal Validity Threats}
Prompt design bias arises because the structure, clarity, and phrasing of prompts may independently influence model performance. Although consistent prompt templates were used and reviewed for clarity, subtle differences in prompt formulation could still affect outcomes. Future work should consider systematically varying prompts to better understand and mitigate this bias. Furthermore, example selection bias affects prompting techniques that rely on examples to illustrate tasks, as the chosen examples may favor certain task types, coding styles, or reasoning approaches, potentially skewing model generalization. While we aimed to select representative examples aligned with the domain, expanding examples' diversity or employing randomized sampling could improve robustness in subsequent studies.

\subsubsection{External Validity Threats}
Various threats to external validity limit the generalizability of our findings. Primarily, model dependence is a concern, as most experiments were conducted using a single or closely related family of LLMs. Since models vary significantly, our results may not transfer to other LLMs like GPT-4 or CodeLLaMa. Similarly, task dependence is a threat, as all code generation tasks were specific to the ASML domain, reflecting its unique software patterns and constraints. This specialization may limit the applicability of our findings to other industries. Lastly, a constant prompting strategy was used across all models and experiments, but different models may perform better with distinct prompting approaches. Future research should explore a broader range of models, tasks, and adaptive prompting strategies to enhance the generalizability of the findings.

\subsubsection{Construct Validity Threats}
We identified several threats to construct validity across the experiments conducted in this study. A key limitation is sparse unit test coverage, as many generated code artifacts lacked sufficient tests, restricting comprehensive verification of correctness; although manual inspection partially mitigated this, more extensive automated test suites would strengthen confidence in the results. Additionally, many of our metrics are binary indicators (e.g., pass/fail), which fail to capture partial correctness or subtle errors. Lastly, metric interdependence is a concern because certain metrics (e.g., unit test results, code quality) are only assessed on code that first builds successfully, introducing a filtering effect that biases downstream evaluations; designing evaluation pipelines to decouple or better account for these dependencies would improve the reliability of findings

\section{Future Work}
Several promising directions arise naturally from this work. First, the evaluation of generated code could be deepened by introducing human-in-the-loop assessments, where domain experts judge qualitative attributes such as readability, maintainability, and security. Complementary developer user studies would provide practical insights into usability, including productivity gains and levels of trust in generated code during real-world tasks.

The evaluation framework itself could be extended along two dimensions: (1) broadening unit test coverage across all garages and (2) incorporating system-level and integration testing to capture inter-component behavior and holistic functional correctness. On the modeling side, parameter-efficient fine-tuning techniques (e.g., LoRA) present an intriguing avenue for leveraging domain-specific data to determine whether measurable improvements can be achieved without the cost of full model retraining.

Another direction is the integration of agentic pipelines~\cite{ionescu2025multi}, enabling iterative, self-correcting workflows that could enhance reliability in domain-specific code generation. Similarly, the context retrieval strategy could benefit from retrieval-augmented generation, in which contextually relevant files are surfaced based on semantic similarity, potentially improving both precision and coherence of generated solutions.

Collectively, these directions offer a clear path to bridge research prototypes and industrial deployment of LLM-based code generation. Pursuing them requires dedicated experiments, user studies, and infrastructure that extend beyond the scope and objectives of our present collaboration, and we therefore leave them to future work.

\section{Conclusion}
We investigated the use of LLMs for domain-specific code generation in the ASML leveling domain, examining three key aspects: prompting strategies, model specialization, and model size. First, we demonstrated that providing examples in prompts substantially improves code quality. Few-shot and one-shot chain-of-thought prompting consistently outperformed zero-shot prompting, highlighting the importance of guiding models with illustrative examples. Second, code-specific models generally surpassed their generic counterparts, with the most pronounced gains observed in the Gemma and DeepSeek-V2 families. Third, while increasing model size improved performance, the benefits tapered off at the higher end, raising important considerations about the cost–performance trade-off.

In all, despite the progress enabled by LLMs, ensuring functional correctness remains a key challenge in part due to limited unit test coverage and the tendency of smaller models to generate non-functional boilerplate. While we introduced build@k as a step toward addressing this challenge, tackling these limitations is essential for realizing reliable, domain-specific code generation in industrial settings.

% \newpage
\bibliographystyle{IEEEtran.bst}
\bibliography{main}
\end{document}